Title

# Disruption of Saturn's ring particles by thermal stress

## Authors
Naoyuki Hirata[1, *], Ryuji Morishima[2], Keiji Ohtsuki[1], Akiko M. Nakamura[1]

## Affiliation
[1] Graduate School of Science, Kobe University, Rokkodai, Kobe 657-8501, Japan
[2] WOTA Corporation, Kitaohtsuka, Tokyo 170-0004, Japan
* Corresponding Author Email address: hirata@tiger.kobe-u.ac.jp

Editorial Correspondence to:
Dr. Naoyuki Hirata
Kobe University
Rokkodai 1-1 657-0013
Tel/Fax +81-7-8803-6566
Email: hirata@tiger.kobe-u.ac.jp

### Highlights
■ We calculated thermal stress acting on Saturn's ring particles
■ We demonstrate that the stress can play a major role in the breakdown of the ring particles
■  A model is provided to explain the lack of particles of >10 m in Saturn's ring

Abstract

Spacecraft and ground-based observations show that the main rings of Saturn lack particles larger than 10 m. Tidal or collisional destruction of satellites/comets have been proposed as the origin of the main rings; however, Saturn's tide alone cannot grind km-sized fragments into submeter-sized particles because of the high mechanical strength of water ice and rock. The question arises as to why such large particles are not left in the current ring. It is known that thermal stress induced by diurnal and seasonal temperature variations can cause weathering and fragmentation of boulders and contribute to dust and regolith production on the Moon and terrestrial planets, and then such thermal stress can break particles larger than a critical radius while cannot smaller than the critical radius. In this study, we examined the role of thermal stress acting on Saturn's ring particles. We found that thermal stress can grind porous ring particles larger than 10–20 m, which explains the lack of particles larger than 10 m in Saturn's ring. Also, fragmentation by thermal stress can be adoptable for the ε rings of Uranus. Furthermore, thermal stress caused by diurnal or seasonal temperature variation acting on boulders on surfaces of icy satellites and asteroids may play an important role in the evolution of their sizes. Our calculations explain the lack of boulders on icy satellites, except in the geologically active provinces such as the tiger stripes of Enceladus, where boulders are supplied by recent geological activity. We predict that future observations can find numerous boulders around Europa's geologically active cracks.

# 1. Introduction

The main rings of Saturn lack particles larger than 10 m. Forward-scattered signals from radio or stellar occultation indicate that the main rings of Saturn mostly consist of particles ranging in size from 1 cm to few meters in radius. The differential size frequency distribution of ring particles follows a power law with an exponent of $-3$ for radii between 1 cm and several meters, whereas it follows a power law with an exponent of $-5$ to $-11$ for radii of $> 10$ m (Marouf et al., 1983; Zebker et al., 1985; French and Nicholson, 2000). Particles larger than 10 m in the main rings are often called moonlets, most of which are estimated to range from approximately 40 to 860 m in diameter (Tiscareno et al., 2006; Seiß et al., 2019). These moonlets were likely produced by the fragmentation of the progenitor moon (Sremčević et al., 2007). The larger objects, such as Pan and Daphnis, are considered rubble pile moons, consisting of a dense core covered with porous regolith layers that originated either from the ring particles (Porco et al., 2007) or through the collision of similar-sized moons (Leleu et al., 2018).

The origin of the deficiency of particles larger than 10 m in the main rings is still under debate. The tidal destruction of satellites, comets, or asteroids has been proposed as the origin of Saturn's main rings (Pollack et al., 1976; Harris, 1984; Dones, 1991; Dones et al., 2007; Charnoz et al., 2009; Canup, 2010; Hyodo et al., 2017). Although Saturn's tide can destroy a 200-km object in its Roche zone, it is unable to grind km-sized or smaller fragments into submeter-sized particles because of the mechanical strength of water ice and rock (Pollack et al., 1973; Pollack, 1975; Goldreich and Tremaine, 1982; Davidsson, 1999). It is rather strange that such large fragments are not left in the current rings. It was proposed that catastrophic events, induced by heliocentric impactors, can reduce the size of the large fragments (Pollack et al., 1973; Pollack, 1975). The collisional cascade model proposed that external bombardment by heliocentric impactors leads to catastrophic fragmentation of small moons around the outer planets, and these fragments further grind the larger fragments into smaller particles (Colwell and Esposito, 1992; Colwell et al., 2000). Hyodo et al. (2017) proposed that the highly eccentric orbits of tidally split fragments, originating from an encountered asteroid, can generate high-energy impacts between the fragments, which can grind the initial fragments into small particles.

Numerous studies have addressed the generation or destruction of ring particles by processes such as tidal disruptions, mutual collisions, gravitational attraction, and adhesions (e.g. Charnoz et al., 2009). However, no study has focused on the effect of thermal stress on Saturn's ring particles. On the Moon, the Earth, other terrestrial bodies, and asteroids, thermal stress has been considered as an important mechanism for weathering and fragmentation of boulders on their surfaces. Boulders exposed to periodic diurnal and seasonal temperature variations develop enough thermal stress to induce thermal fatigue crack growth (e.g. McFadden et al., 2005; Viles et al., 2010; Delbo et al., 2014; Basilevsky et al., 2015; Molaro et al., 2015; Molaro et al., 2017 Hazeli et al. 2018; Ravaji et al. 2019; El Mir et al. 2019; Liang et al. 2020; Uribe-Suarez et al. 2021). For example, the daytime and the nighttime temperatures near the lunar equator reach 400 K and 140 K, respectively, which generate thermal stress on the order of 10 MPa in meter-sized boulders (Molaro et al., 2017). The survival time (i.e., the time required to break boulders by thermal fatigue) of meter-sized boulders on the Moon is estimated to be < $10^4$ years (Delbo et al., 2014). Considering this, we expect that thermal stress can reduce the size of the large fragments.

In the present work, we examined the role of thermal stress on the evolution of particle sizes in Saturn's rings. We provide an overview of the observed temperature variations of Saturn's ring particles in Section 2; estimate the physical properties of the ring particles such as porosity, thermal conductivity, and mechanical strength in Section 3; calculate the skin depth and thermal stress using a simple approximation in Section 4; calculate them more adequately using the one-dimensional non-steady state spherical thermal diffusion equation in Section 5; and discuss implications of the present study for the particle size evolution of Saturn's rings and Uranus' rings as well as for the evolution of ice boulders on the surfaces of Jovian and Saturnian satellites in Section 6.

## 2. Observed temperature variations of Saturn's ring particles

Temperature variations of Saturn's ring particles are complicated, they are inferred from the observational data by Composite Infrared Spectrometer (CIRS) onboard the Cassini spacecraft, as summarized below (Spilker et al., 2006; Ferrari and Reffet, 2013; Spilker et al., 2013).

1. The ring-opening angle relative to the Sun ($B$) causes seasonal temperature variations in the ring particles (Fig. 1a). Saturn's equator and the ring plane are tilted relative to its orbit around the Sun by 27°. When the opening angle reaches the largest value ($B = 27°$), that is, when Saturn is at solstice, the ring plane receives the maximum solar radiation flux, and the ring temperature becomes the highest. Then, the sunlit face (i.e., the side facing the sun) of the ring plane reaches 90 K, and the unlit face (i.e., the opposite side) reaches 65 K. When the opening angle reaches the smallest value ($B = 0°$), that is, when Saturn is at equinox, the ring plane receives limited solar radiation flux, and the ring temperature becomes the lowest. Then, the ring plane receives only Saturn flux (i.e., Saturnshine and thermal emission from Saturn), and its temperature cools down to 50 K. This seasonal variation has a period of half Saturn year (one Saturn year = 29.5 Earth years) because the solstice and equinox occur twice a year.

    2. The rotation of each ring particle causes diurnal temperature variations in the ring particles (Fig. 1b). Each ring particle has a dayside and nightside, and the rotation of the particles causes periodic temperature variation on its surface. Small ring particles (<10 cm) spin ten to hundred times, large particles (1–10 m) spin few times, and moonlets (> 40 m) spin only once in one orbital period around Saturn (Ohtsuki, 2005; Morishima and Salo, 2006). Note that one orbital period of a typical ring particle around Saturn is ~10 h. Note that the polar regions of large moonlets have polar nights because they are tidally locked with the planet, and therefore, the seasonal variation may cause an asymmetric thermal structure on their surfaces.

    3. The passage of the shadow of Saturn causes periodic cooling of ring particles (Fig. 1c). These temperature variations depend on the ring-opening angle. The observed temperature offset between the ingress and egress of Saturn's shadow is approximately 10 K when $B = 27°$, whereas it is negligibly small when $B = 0°$. The passage of Saturn's shadow lasts approximately 2 h in each orbital period.

    4. In an optically thin ring, the constituting particles frequently moves between the sunlit and unlit faces (Fig. 1d). When a particle is near the sunlit face, it gets heated up, and when the particle is near the unlit face, it cools down. In contrast, ring particles barely move between the sunlit and unlit faces in an optically thick ring (Morishima and Salo, 2006). In the latter case (the A or B rings), the observed temperature offset between the sunlit and unlit faces of the ring plane is 25 K when $B = 27°$, whereas it is negligibly small when $B = 0°$. In case of optically thinner rings (such as the C ring), the

temperature difference between the sunlit and unlit faces is small even when $B$ = 27°. Although a precise estimate of the periods of such a type of temperature change is difficult, they are roughly on the order of one orbital period.

As the second, third, and fourth periodic variations described above are short-term with duration of < 10 h, their thermal skin depths are extremely small (see Section 4). Therefore, the associated temperature changes on particle surfaces are relatively surficial (with a corresponding skin depth on the order of 1 cm). However, the seasonal variations are long-term, and the associated temperature changes on particle surfaces penetrate to its interior regions (on the order of 1 m to 10 m). Because the rotation period of ring particles is considerably shorter than half Saturn year, each particle would have spherically symmetric thermal structure, except for its surface.

## 3. Physical properties of ring particles

Physical properties of water ice depend on temperature, composition of components, internal structure, and porosity. Especially, porosity has a significant impact on the thermal and mechanical properties. Although properties of icy particles in the main rings are poorly constrained, for simplicity, we consider three types of water ice ring particles according to their porosity, as we describe below. Also, we assume that ring particles are sufficiently rigid. If ring particles are loose aggregates that have a structure that allows dissipation of stress generated by thermal volume change, thermal stress would hardly act on such aggregates as a whole (see also 8th paragraph in Discussion).

### 3.1. Thermal conductivity

Thermal properties of icy particles, such as their conductivity, vary depending on the temperature, composition of components, internal structure, and the porosity of ice. Especially, the conductivity is dominantly affected by the porosity (Ross and Kargel, 1998); for example, the thermal conductivity of water ice is reduced by several orders of magnitude depending on the porosity (Krause et al., 2011). Porosity of icy satellites can be reduced over time due to various structural evolutions induced by various mechanisms including creeping, sintering, melting, impact, and differentiation, although it is expected to still remain at a certain degree (Matson et al., 2009). Laboratory measurements by Durham et al. (2005) demonstrated that in pure water, 20% porosity can be sustained at pressure as high as 150 MPa when the temperature is < 120 K. Note that the central pressure in Mimas is 7.4 MPa (Matson et al., 2009). Considering the bulk densities of small Saturnian moons and Hyperion, it is reasonable to assume that icy satellites have porosity of 40% (Matson et al., 2009). When ring particles originate from disruption of such mid-sized icy objects, they would also have initial porosity of 40 %. However, when ring particles originate from a well-compacted icy mantle of Titan-sized object, as proposed by Canup (2010), they can have initial porosity approximately 0%. The observed thermal inertia of ring particles indicates that the particles are covered with highly porous regolith layers with a porosity of 90%, and its conductivity is as low as $k$ = 0.0022 Wm$^{-1}$K$^{-1}$ (Morishima et al. 2011; 2014).

In the case of porous ice, physical binding at inter-particle contacts (i.e., sintering) also affects various properties of ice; the growth of the neck area of the inter-particle contacts can increase thermal conductivity and mechanical strength (Gundlach et al., 2018). Because of the high ring temperature at the solstice (100 – 120 K), it is possible that sintering occurs in dust grains of the main rings. For example, 50% porosity water ice composing of 5-µm grains has a thermal conductivity of $k$ = 0.3 Wm$^{-1}$K$^{-1}$ when the neck area reaches the maximum value, whereas it has thermal conductivity of $k$ = 0.08 Wm$^{-1}$K$^{-1}$ when sintering does not occur at all (Gundlach et al., 2018).

Based on the above views, for simplicity, we consider three types of water ice ring particles: i) non-porous particles ($\phi$ =0% porosity) with a thermal conductivity of $k$ = 12 Wm$^{-1}$K$^{-1}$, ii) porous particles ($\phi$ =40%) with $k$ = 0.10 Wm$^{-1}$K$^{-1}$, and iii) highly porous particles ($\phi$ =90%) with $k$ = 0.0022 Wm$^{-1}$K$^{-1}$ (Table 1). We assume that the heat capacity of water ice is constant, $C$ = 450 J kg$^{-1}$K$^{-1}$ (Shulman, 2004).

### 3.2. Young's modulus and tensile strength

Mechanical properties of porous icy particles, such as Young's modulus and tensile strength, also largely depend on porosity. Polycrystalline granular ice, which is close to ideal non-porous ice, has Young's modulus of $E$ = 9.3 GPa at 263 K (Schulson and Duval, 2009). The relationship between Young's modulus and the density of snow is approximated by $E_{snow} = (\rho_{snow}/\rho_{ice})^2 E_{ice}$ (Gibson and Ashby, 1988). Yasui and Arakawa (2010) obtained a relationship $E = 9.1 - 18.9\phi$ in the range of $0 \leq \phi \leq 20\%$. Following Gibson and Ashby (1988), we consider $E$ = 9.3 GPa, 3.3 GPa, and 0.093 GPa for non-porous, porous, and highly porous particles, respectively. Young's modulus does not significantly depend on temperature (Mellor, 1975). We assume a constant Poisson ratio of $v$ = 0.32 (Schulson and Duval, 2009).

Strength of ice decreases with increasing porosity; for example, Mellor (1975) obtained a power-law relation for the porosity dependence of the compressive strength of pure water ice, as

$$Y \sim Y_0 \left(1 - \frac{\phi}{100\ \%}\right)^n, \quad (1)$$

where $n$ is a scaling parameter, and $Y_0$ is the strength of the non-porous ice. Mellor (1975) obtained $n$ = 4.2 in the density range from 150 to 900 kg/m$^3$. Yasui and Arakawa (2010) obtained the value of $Y_0$ in pure water ice at 253 K with $n$ = 3–4 in the range of $0 \leq \phi \leq 20\%$. Note that the degree of sintering affects the tensile strength of porous ice. For example, 50% porosity water ice composing of 5-µm grains has tensile strength of $Y$ = 1.0 MPa when the neck area reaches the maximum, and the value of $Y$ becomes 0.01 MPa when sintering does not take place (Gundlach et al., 2018). The tensile strength of non-porous pure water ice varies from 0.7 to 3.1 MPa, depending on the temperature, strain rate, and minor components (Petrovic, 2003). Although the mechanical properties of porous ice and the degree of sintering in the main rings are poorly constrained, in this study, adopting $n$ = 4 and $Y_0$ = 1 MPa in Eq. (1), here we assume $Y$ = 1.0 MPa, 0.13 MPa, and 0.00010 MPa for non-porous, porous, and highly porous particles, respectively. Values of the parameters adopted in each particle model are summarized in Table 1.

## 4. Simple approximation for thermal stress acting on ring particles

Boulders on the surfaces of airless bodies in the terrestrial region whose size is several times the skin depth are generally most susceptible to breakdown by thermal stress (Molaro et al. 2017). Skin depth is a measure of the depth at which the amplitude of the periodic temperature variation is reduced to $1/e$ times its value at the surface. The theoretical skin depth in the semi-infinite half-space is written by (Turcotte and Schubert, 2001):

$$d_\omega = \sqrt{\frac{2k}{\rho C \omega}}, \quad (2)$$

where $k$ is the thermal conductivity, $C$ is the specific heat capacity, $\rho$ is the density of the target, and $\omega$ is the frequency of temperature change at the surface ($\omega = 2\pi/T_p$, where $T_p$ is the period of the periodic temperature change) (Fig. 2). When the radius of a boulder is sufficiently larger than the skin depth, the temperature in the boulder sub-surface (i.e., within the skin depth) varies periodically, while the sufficiently deep interior (i.e., below several times the skin depth) becomes isothermal (Fig. 3A). Such an inhomogeneity in temperature profile causes thermal expansion/contraction of localized areas, which leads to strong thermal stress on the surface. When the radius of a boulder is sufficiently smaller than the skin depth, the temperature profile is always homogenous throughout the boulder, which causes uniform thermal expansion/contraction, and does not lead to strong thermal stress (Fig. 3B). On the other hand, when the radius of a boulder is close to several times the skin depth, the period of the periodic temperature variations in the center gets halved from that in the surface, which leads to strong thermal stress throughout the boulder (Fig. 3C). Therefore, ring particles whose size is several times the skin depth are most susceptible to breakdown by thermal stress. For example, the skin depth of non-porous particles was $d_\omega = 67$ m for the 15-year seasonal variation (Table 1). This indicates that 100-m sized non-porous ice particles around Saturn are most affected by thermal stress, whereas sub-decameter-sized particles are only slightly affected by seasonal thermal stress.

The thermal stress acting on the particle surface is approximated by
$$\sigma \sim E\alpha_l \, \Delta T, \quad (3)$$
where $\Delta T$ is the temperature difference between the particle surface and the particle center, $E$ is Young's modulus, and $\alpha_l$ is the coefficient of linear thermal expansion of the material (Timoshenk and Goodier, 1970). More accurate equations for stress components are given in Eq. (12) and (13). Furthermore, $\alpha_l$ also depends on temperature, which can be approximated by a simple linear function of temperature between 30 K and 250 K (Rottger et al., 1994):
$$\alpha_l(T) = 2.82 \times 10^{-7} T - 2.06 \times 10^{-5} \text{ K}^{-1}. \quad (4)$$
Due to the lack of experiments, it is unknown whether the coefficient of linear thermal expansion of ice is affected by porosity or not. However, it is established that the coefficient for porous alumina structures is independent of porosity (Hirata et al., 2017). Therefore, we assume that $\alpha_l$ for water ice is also independent of porosity. For example, when there is a temperature difference of $\Delta T = 10$ K between the center and the surface, we obtain $\sigma \sim 0.71$ MPa for non-porous particle, $\sigma \sim 0.25$ MPa for porous particle, and $\sigma \sim 7$ kPa for highly porous particle, respectively. These values of thermal stress are comparable to or greater than the estimated tensile strength shown in Table 1. In addition, these values are considerably greater than the tidal stresses, $\rho R^2 \omega_{\text{orb}}^2$, where $\omega_{\text{orb}}$ is the orbital angular velocity of the particle (Weidenschilling et al., 1984).

The minimum temperature difference required to induce thermal stress sufficient to overcome the tensile strength can be written as

$$\Delta T \sim \frac{Y_0(1-\phi)^n}{E\alpha_l} . \quad (5)$$

When $T = 100$ K is considered in the initial condition, we can obtain $\Delta T = 14$ K for non-porous particle, $\Delta T = 5.2$ K for porous particle, and $\Delta T = 0.14$ K for highly porous particle, respectively.

## 5. Calculation of thermal stress acting on ring particles

We obtained the temperature profile of a ring particle following the seasonal temperature variation model developed by Morishima et al. (2016). In this section, we briefly describe the model and the validity of each input parameter. The other types of temperature variations (described in Fig. 1b, c, d) might also lead to sufficient thermal stress to overcome the tensile strength (Appendix A), however, the corresponding skin depth in such cases is expected to be too small to influence ring particle size evolution (although it must contribute dust and regolith production on the main rings); therefore, here we focus on the seasonal temperature variation. In this model, it is assumed that ring particles are homogenous spheres with a radius of $R$. Because the rotation period as well as the orbital period of ring particles is considerably shorter than a half Saturn year, temperature variations due to rotation or passage of Saturn's shadow are homogenized in the interior of the particles. Therefore, the seasonal temperature profile of ring particles is expected to be spherically symmetric; the profile is described by a function of the radial distance from the particle center.

The time evolution of the particle temperature, $T$, can be obtained by solving the following one-dimensional non-steady-state spherical thermal diffusion equation:

$$\frac{\partial T}{\partial t} = \frac{k}{\rho C r^2} \frac{\partial}{\partial r}\left(r^2 \frac{\partial T}{\partial r}\right), \quad (6)$$

where $r$ is the radial distance from the center of the particle. We numerically solved Eq. (6) using the Crank–Nicholson method (Press et al., 1986). The boundary condition at the surface is given as follows:

$$k \frac{\partial T}{\partial r} = F_{\text{sun}} + F_{\text{Saturn}} - \epsilon \sigma T^4 \left(1 - \frac{\Omega_p}{4\pi}\right), \quad (7)$$

where $F_{sun}$ and $F_{saturn}$ are the fluxes from the Sun and Saturn, respectively; $\epsilon$ is the infrared emissivity (we consider $\epsilon = 1$); $\sigma$ is the Stefan-Boltzmann constant; and $\Omega_p$ is the solid angle of the surrounding particles that depends on the ring optical depth ($\tau$) according to the equation $\Omega_p = 6\{1 - \exp(-\tau)\}$ (Ferrari et al., 2005). For simplicity, we assume $\tau = 0.5$ for the A ring, $\tau = 0.7$ for the B ring, and $\tau = 0.1$ for the C ring, respectively (Esposito et al., 1983). The solar flux is given as follows:

$$F_{\text{sun}} = \frac{1-A}{f} S(B,\tau) \left(\frac{1 \text{ AU}}{a_{saturn}}\right)^2 F_\odot, \quad (8)$$

where $A$ is the albedo of the particle ($A_{lit}$ for the sunlit face and $A_{unlit}$ for the unlit face), $f$ is the spin factor, $F_\odot = 1370$ W m$^{-2}$ is the solar flux at 1AU, $S$ is the factor describing the shadowing of surrounding particles, and $a_{saturn}$ is the orbital distance of Saturn from the Sun. We consider $f = 3$ (Spilker et al., 2013), $A_{lit} = 0.498$, and $A_{unlit} = 0.625$ (Morishima et al., 2016). The shadowing factor, $S$, is approximated as follows:

$$S(B,\tau) = \frac{\sin|B|}{1-\exp(-\tau)}\left[1 - \exp\left(-\frac{\tau}{\sin|B|}\right)\right], \quad (9)$$

where $B$ is the ring-opening angle with respect to the Sun (Altobeli et al., 2008). Fig. 4 demonstrates the relationships among $S$, $B$, and $\tau$. When the ring optical depth is sufficiently low, $S$ increases exponentially near Saturn's equinox. This means that the temperature change is more dramatically in an optically thinner ring. The Saturn flux is given as follows:

$$F_{\text{saturn}} = \epsilon \sigma T_{eq}^4 \left(1 - \frac{\Omega_p}{4\pi}\right), \quad (10)$$

where $T_{eq} = 43$ K is the equilibrium equinox temperature exclusively due to Saturn's flux. The boundary condition at the center of the particle is given as follows:

$$\frac{\Delta r}{3} \rho C \frac{\partial T}{\partial t} = k \left.\frac{\partial T}{\partial r}\right|_{r=\Delta r}. \quad (11)$$

This equation describes heat balance of a sphere with radius Δr at the particle center. When the interior of the particle is divided into 100 layers with an identical thickness, then Δr = R/100.

In the case of spherically symmetrical temperature distribution with respect to the center of the particle, the thermal stress, consisting of a radial stress component ($\sigma_r$) and two tangential stress components ($\sigma_t$), is also described by a function of r. Then, the stress components are written by (Timoshenk and Goodier, 1970)

$$\sigma_r(r) = \frac{2E}{1-v}\left(\frac{1}{R^3}\int_0^R \alpha_l T(r) r^2 dr - \frac{1}{r^3}\int_0^r \alpha_l T(r) r^2 dr\right), \text{ and} \quad (12)$$

$$\sigma_t(r) = \frac{E}{1-v}\left(\frac{2}{R^3}\int_0^R \alpha_l T(r) r^2 dr + \frac{1}{r^3}\int_0^r \alpha_l T(r) r^2 dr - \alpha_l T(r)\right). \quad (13)$$

Using these equations, we can obtain the temperature and thermal stress profiles of the ring particles.

The temperature and thermal stress profiles of non-porous, porous, and highly porous particles in the B ring are provided in Figs 5, 6, and 7, and in Movies S1, S2, and S3, respectively. For the highly porous particles, the temperature variation is ΔT = 30 K, the skin depth is ~ 5 m, and the maximum thermal stress at the surface is σ ~ 30 kPa. For the porous particles, the temperature variation is ΔT = 20 K, the skin depth is ~ 20 m, and maximum thermal stress on the surface is σ ~ 500 kPa. For the non-porous particles, the temperature variation is ΔT = 10–30 K, the skin depth is approximately 100 m, and maximum thermal stress on the surface is σ ~ 0.8 MPa. The skin depth and the relationship between the stress and temperature variations are consistent with the simple approximation shown in Eq. (2) and (3), respectively.

Fig. 8 shows the plots of the maximum tangential component of the tensile stress acting at the particle surface or at its center during one Saturn year as a function of the particle size. The stress at the particle surface is always greater than that at the center because the temperature variation at the surface is larger than that at the center. We have considered only the tangential component of tensile stress for the following reasons: (1) the radial component of thermal stress corresponds to the tangential component at the particle center, (2) the radial component of thermal stress is always zero at the particle surface, and (3) the compressive strength of water ice is usually ten times stronger than the tensile stress, and tensile stress is much easier to cause the breakdown of particles than compressive stress. We can interpret that (i) when the stress at the particle center overcomes the particle strength, the entire particle will be cracked, (ii) when the stress at the particle surface (and not at the center) overcomes the particle strength, only the surface of the particle will get cracked, and (iii) when the stress at the particle surface does not overcome the particle strength, the particle will not get cracked anywhere.

The radius of particles that experience the strongest tensile stress is 10 m for highly porous particles, 20 m for porous particles, and 200 m for non-porous particles, respectively (Fig. 8). These radii corresponding to the peak tensile stress are roughly equal to several times the skin depth, as shown in Eq. (2). The peak radius and shape of the curves hardly depend on the optical depth of the rings. In the case of highly porous particles (bottom panel), even small particles with < 1m experience stronger thermal stress compared to the tensile strength. In the case of porous particles (middle panel), several meters or larger particles experience thermal stress larger than the tensile strength. On the other hand, non-porous particles do not experience any thermal stress larger than the tensile strength (top panel) because the temperature variation created within the particle is not sufficiently large in this case.

## 6. Discussion

Our results show that thermal stress causes the breakdown of particles that are larger than a critical radius; for example, it is 10 meter in the case of ice particles with 40% porosity (Fig. 8). It should be noted that, although this critical radius is determined by the porosity, the coefficient of linear thermal expansion, Young's modulus, and tensile strength of ice, there is little experimental work on the rheology of ice at 50–100 K. Therefore, further experimental work is necessary to improve the understanding of the role of thermal stress on icy bodies in the outer solar system. In addition, we did not consider the role of thermal fatigue. It is also possible that thermal fatigue, rather than thermal shock, plays a role in the failure of ring particles to achieve a larger size. Even if the thermal stress is weaker than the tensile strength, the periodic thermal stress can break up the particles. We demonstrate that non-porous particles do not experience thermal stress larger than the tensile strength; however, the created thermal stress exceeds 0.8 MPa. It has been shown that the critical shear stress for the onset of crack nucleation is 0.8 MPa (Schulson et al., 1984; Cole, 1988; Hammond et al.,

2018); therefore, thermal fatigue may occur in non-porous particles. Even if ring particles originate from a well-compacted mantle of a Titan-sized object and have very low porosity approximately 0 %, the thermal stress could have played a role in grinding large fragments into meter-sized particles. Nonetheless, there are many unknown aspects of thermal fatigue in the case of water ice samples; for example, as far as we know, no experiments for thermal fatigue in relation to the evolution of icy bodies in the outer solar system have been performed for ice samples at 50–100 K. Experiments performed by Hammond et al. (2018) demonstrated that thermal fatigue does not occur for ice samples at 198 K. Their results cannot be directly applied for ice samples at 50–100 K. Further experiments for ice at low temperatures are necessary for a better understanding of thermal fatigue of water ice and the lifetime of ring particles.

Cassini observations suggest that there are 10 m or larger objects, such as moonlets, in the main rings. Diurnal and seasonal thermal stress cause cracking at the object surfaces, which can generate a porous regolith layer on the objects. Such a porous regolith layer, which plays a role in heat insulator, can keep the object's interior isothermal, thereby prolonging the lifetime of the entire object. Moreover, such a regolith layer may not be a stable structure for objects in the main ring region because of tidal stripping, collision, and/or weak gravitational interaction with other particles, and therefore, its thickness would become thinner or thicker frequently and a rigid part of the moonlets is not always isothermal. Regolith layers can be occasionally released from the object, which can become freely-floating new ring particles, thereby gradually diminishing the object size. It is possible that the moonlets are transient objects of such processes.

Based on the present results, we propose a scenario for the size evolution of particles in the rings. After the tidal destruction of a progenitor satellite, comet, or asteroid, numerous decameter-sized or larger fragments form a debris disk around Saturn or moonlets in the main rings. Thermal stress would have further ground such fragments into smaller particles. Thermal stress would also create regolith layers on the surface of the fragments, and the regolith particles were released occasionally from the fragments, which became new freely floating ring particles, and the fragments gradually became smaller.

It is known that millimeter-sized or smaller particles are lacking in

Saturn's main rings (Marouf et al., 1983; Zebker et al., 1985; French and Nicholson, 2000). Thermal stress due to temperature variations in shallow skin depths induced by particle spin or the passages of Saturn's shadow (shown in Fig. 1b,c,d) should also generate dusty ring particles. Even if sub-centimeter particles are created, they would likely stick to other larger ring particles due to cohesion, as proposed by previous studies explaining the observed lower size cutoff of Saturn's ring particles (Dones et al., 1993, Bodrova et al., 2012; Ohtsuki et al., 2020). This view is consistent with (i) spectral reflectance studies in ultraviolet, visible, and infrared wavelength ranges, reporting that ring particles are covered with grains of radius 5–20 µm (Doyle et al., 1989; Poulet et al., 2003; Nicholson et al., 2008; Bradley et al., 2010); and (ii) the measurement of thermal inertia of the main rings, which indicates that ring particles are covered with a dusty regolith layer (Morishima et al., 2011, 2014). For such particles, removal due to the Yarkovsky effect can also be effective (Rubincam, 2006; Vokrouhlický et al., 2007).

The sizes of particles in the ε ring of Uranus, which is the optically thickest ring of Uranus ($\tau > 0.8$) (Tyler et al., 1986), distribute over a narrow range between 0.2 cm to 30 m and small dust grains are lacking (Lane et al., 1986). The small particle sizes of the ε ring can be also explained by the same manner. When the particles of the ε ring have moderate porosity and the period of seasonal temperature variation is half Uranus year (one Uranus year = 84 Earth years), the seasonal thermal skin depth of ring particles can be estimated to be $d_\omega = 10$ m, as shown in Fig. 2. Since the observed upper size limit of the ε ring is approximately equal to several times the above skin depth, its narrow particle size range can also be explained by disruption due to thermal stress.

Recent observations have revealed that some Centaurs and trans-Neptunian objects such as Chariklo (10199) and Haumea (136108) have multiple rings (Braga-Ribas et al., 2014; Ortiz et al., 2017), and Chiron (2060) may also have rings (Ortiz et al., 2015). We predict that if the rings have seasonal temperature variations, the rings would have an upper size cutoff on the order of 10 m. However, other than the main rings of Saturn and the ε ring of Uranus, most of the ring systems we know are very tenuous. Such faint rings do not have seasonal temperature variations, as shown in Fig. 4, and therefore, seasonal thermal stress is not expected to grind large particles.

Thermal stress induced by particle spin or the passage of the shadow of the central body (shown in Fig. 1b,c,d) usually vary in much shorter timescale than the seasonal variation, and the corresponding skin depth is significantly smaller than particle radius. In this case, very thin regolith layers may be created at the surface of large particles. However, they cannot grind km-sized large particles into meter-sized (although km-sized particles can gradually become somewhat smaller). Therefore, we infer that in the case of an optically thick ring, the upper size limit of particles can be determined by the seasonal skin depth (~10 m), whereas in the case of a tenuous ring, the large particles and the released dusty regolith particles would remain in their orbits.

It should be noted that the calculation of thermal stress in our model assumes that ring particles are rigid. Once large fragments are disrupted into small particles and they form loose aggregates that have a structure that allows dissipation of stress generated by thermal volume change, thermal stress would hardly act on such aggregates as a whole. In the scenario of ring formation by destruction of one or multiple large icy satellite(s), ring particles are originated from condensed icy mantle of the progenitor satellite, and therefore, they would likely be rigid at early evolution. Our thermal stress model would be applicable to such early evolution in this scenario. On the other hand, it is plausible that, at present, ring particles are aggregates loosely connected only by the gravity and adhesion, whose size distribution is a result of competing accretion and fragmentation processes (e.g. Weidenschilling et al., 1984). In this case, relative impact speeds and sizes determine whether coagulation, restitution or fragmentation becomes dominant (e.g. Spahn et al., 2004), and the maximum aggregate size in Saturn's main rings created by collision-caused fragmentation and coagulation is estimated to be on the order of a few tens of meters (Brilliantov et al. 2015, Guimarães et al. 2012). This view is also consistent with the upper cutoff of observed size distribution of Saturn's ring particles, but does not explain how the current particle size distribution was formed from kilometer-sized fragments of a progenitor satellite. It is difficult to argue conclusively rheological properties of fragments of a progenitor satellite or ring particles, however, even in the current main rings, the ring temperature reaches 100 – 120 K at solstice, and therefore, it is possible that temperature-driven sintering can take place in ring particles, which could make dust grains be more rigid. Radiation-driven sintering, taking place in saturnian satellites

(Schaible et al. 2017), could also take place in the main rings. In this case, thermal stress may act to some extent even on loose aggregates, although it should be weaker than the case of rigid ice.

Our model of thermal stress on icy particles can be applied to the size evolution of boulders on the surfaces of the saturnian and jovian icy satellites. Such ice boulders are also thermally stressed by diurnal or seasonal temperature variations, which would shorten their lifetime. Martens et al. (2015) examined Cassini's high-resolution images of Enceladus with a resolution of 4–25 m/px and found that all the boulders of Enceladus exist within 20 km of geologically active cracks, so-called tiger stripes, and not elsewhere. No boulders were found on other Saturnian satellites, as far as we investigated. We consider that boulders on icy satellites have a shorter lifetime due to thermal stress, although Martens et al. (2015) proposed other hypothesis for the removal of Enceladus' boulders. The skin depths are $d_\omega =$ 1.0 m for non-porous ice, 0.12 m for porous ice, and 0.043 m for highly porous ice, assuming that diurnal temperature variation on Enceladus occurs in a period of 32.9 h. Daytime and nighttime temperature offset at the equator of Enceladus is $\Delta T$ = 20 K (Howett et al., 2010), which leads to thermal stress of $\sigma \sim 1.4$ MPa for non-porous particles, $\sigma \sim 0.50$ MPa for porous particles, and $\sigma \sim 14$ kPa for highly porous particles, respectively, based on Eq. (2). This stress overcomes the tensile strength of each particle model. On Enceladus, boulders must have been released from the cracks of the south pole very recently, and thermal stress might be currently reducing the size of boulders. Ground-based observations have reported that Europa is geologically active, similarly to Enceladus (Roth et al., 2014). High-resolution images (~10 m/pixel) obtained by the Galileo spacecraft, which cover only a limited surface of Europa, do not show any boulders. However, we predict that future observations will find many boulders around geological active cracks or similar provinces of the Europe and the existence of boulders will be an evidence of recent geological activity.

## 7. Conclusions

We calculated the thermal stress induced by seasonal temperature variations of ring particles. The strongest tensile stress occurs for particles with a radius of 10 m for highly porous particles, 20 m for porous particles, and 200 m for non-porous particles, respectively. These peak radii are roughly equal to several times the skin depth. In the case of highly porous particles, even small objects with < 1m experience thermal stress larger than the tensile strength. In the case of porous particles, several meter-sized or larger particles experience thermal stress larger than the tensile strength. In contrast, the thermal stress in the case of the non-porous particles is not larger than the tensile strength. However, the results are uncertain because there are only few experimental works on the rheology of water ice at 50–100 K; therefore, further experimental work is necessary for improving our understanding of the role of thermal stress in the outer solar system. Also, we note that our thermal stress model is adoptable only if ring particles are sufficiently rigid. Based on our model, we propose a new scenario for the ring's particle size evolution: (i) After tidal destruction of a satellite, comets, or asteroids, decameter-sized or larger fragments float around Saturn; (ii) thermal stresses further ground km-sized fragments into smaller particles; (iii) thermal stresses induced by seasonal changes of the ring temperature cannot grind sub-meter-sized ring particles. This scenario explains that the main ring lacks particles larger than 10 m. Currently, thermal stress causes the cracking or breakup of ring particles or moonlets, which generates a layer of regolith on large objects in the main rings, and such released regolith particles supply a new freely floating ring particle.

Table 1. Parameters of particles

|  | Non-porous | Porous | Highly porous |
| --- | --- | --- | --- |
| $\phi$ (%) | 0 | 40 | 90 |
| $\rho$ (kgm$^{-3}$) | 933 [*1] | 560 | 100 [*5] |
| $C$ (Jkg$^{-1}$K$^{-1}$) | 450 [*2] | 450 [*2] | 450 [*2] |
| $k$ (Wm$^{-1}$K$^{-1}$) | 12 [*3] | 0.10 [*4] | 0.0022 [*5] |
| $E$ (GPa) | 9.3 [*6] | 3.3 | 0.093 |
| $N$ | 0.32 [*6] | 0.32 [*6] | 0.32 [*6] |
| $Y$ (MPa) | 1.0 [*7] | 0.13 | 0.00010 |
| $d_\omega$ (m) Diurnal (10h) | 0.57 | 0.067 | 0.024 |
| $d_\omega$ (m) Seasonal (15 y) | 67 | 7.7 | 2.7 |

[*1] Feistel and Wagner, (2006)

[*2] Shulman, (2004)

[*3] Andersson and Inaba, (2005)

[*4] Krause et al. (2011)

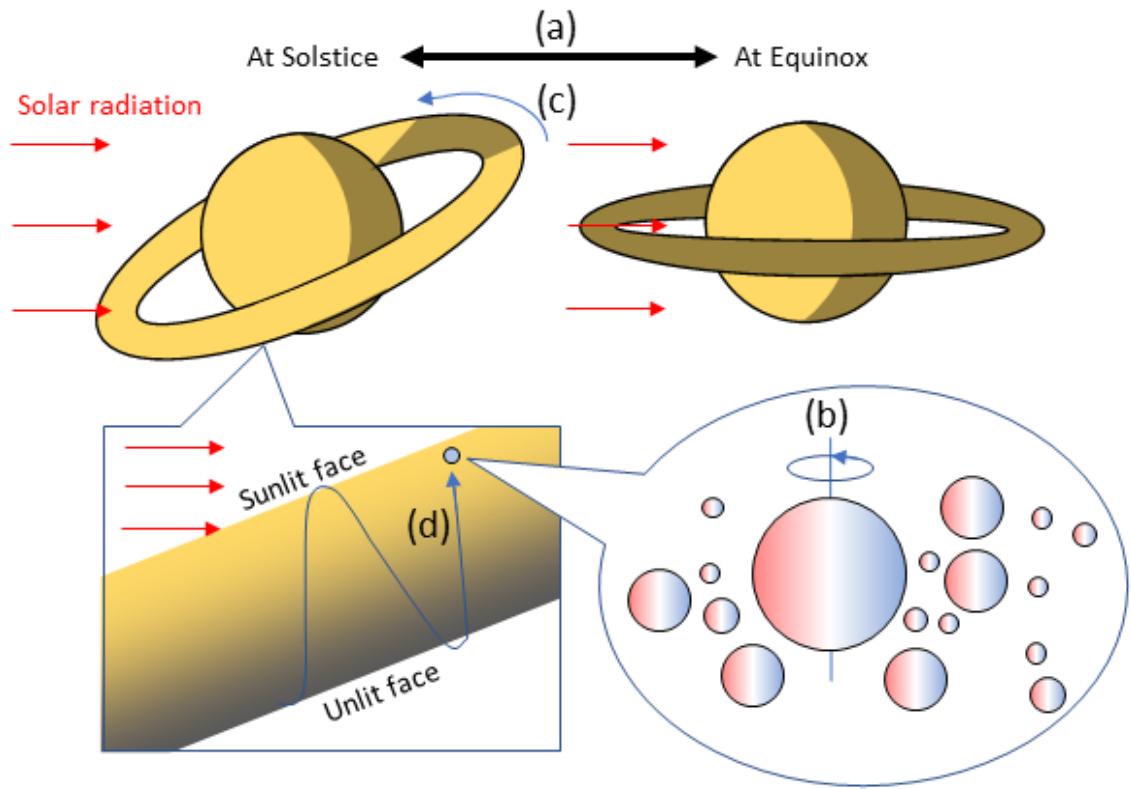

**Figure 1.** Schematic illustration for mechanism for creating periodic temperature variation of ring particles. (a) The ring-opening angle relative to the Sun causes seasonal temperature variation of ring particles. As Saturn approaches its solstice, the ring temperature becomes the highest, and at equinox, the temperature is the lowest. (b) The rotation of each ring particle causes diurnal temperature variation; although surface temperatures of fast rotating particles are homogenous, the dayside and the nightside of the slowly rotating particles are hot and cold, respectively. (c) When a particle comes under Saturn's shadow, it gets cooled down. (d) In an optically thin ring, a particle frequently moves between the sunlit and unlit faces. The particle gets heated up when it is near the sunlit face, and it gets cooled down when in the unlit face.

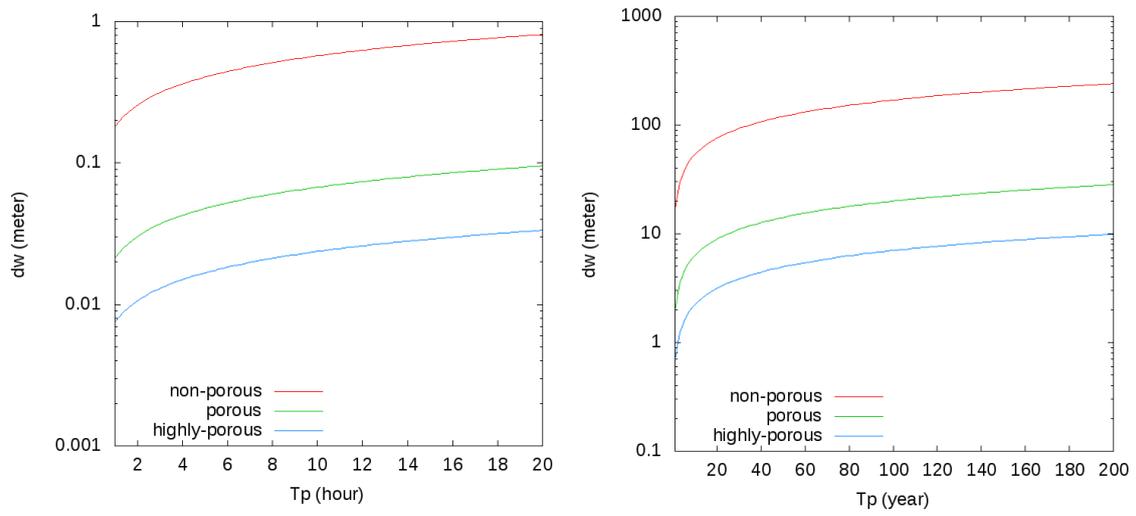

**Figure 2.** Change in thermal skin depth, $d_\omega$ as a function of the period of temperature change, $T_p$ in hours (left) and years (right).

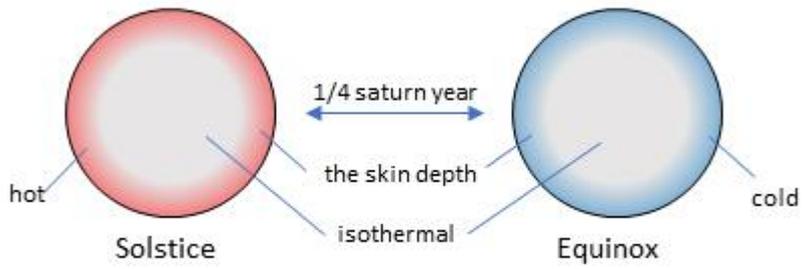
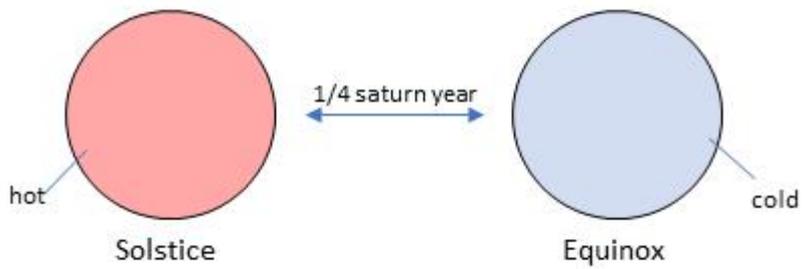
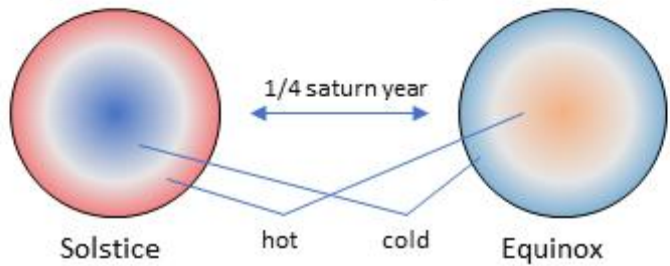

**Figure 3.** Schematic illustration for relationship between temperature profile of a ring particle and its skin depth.

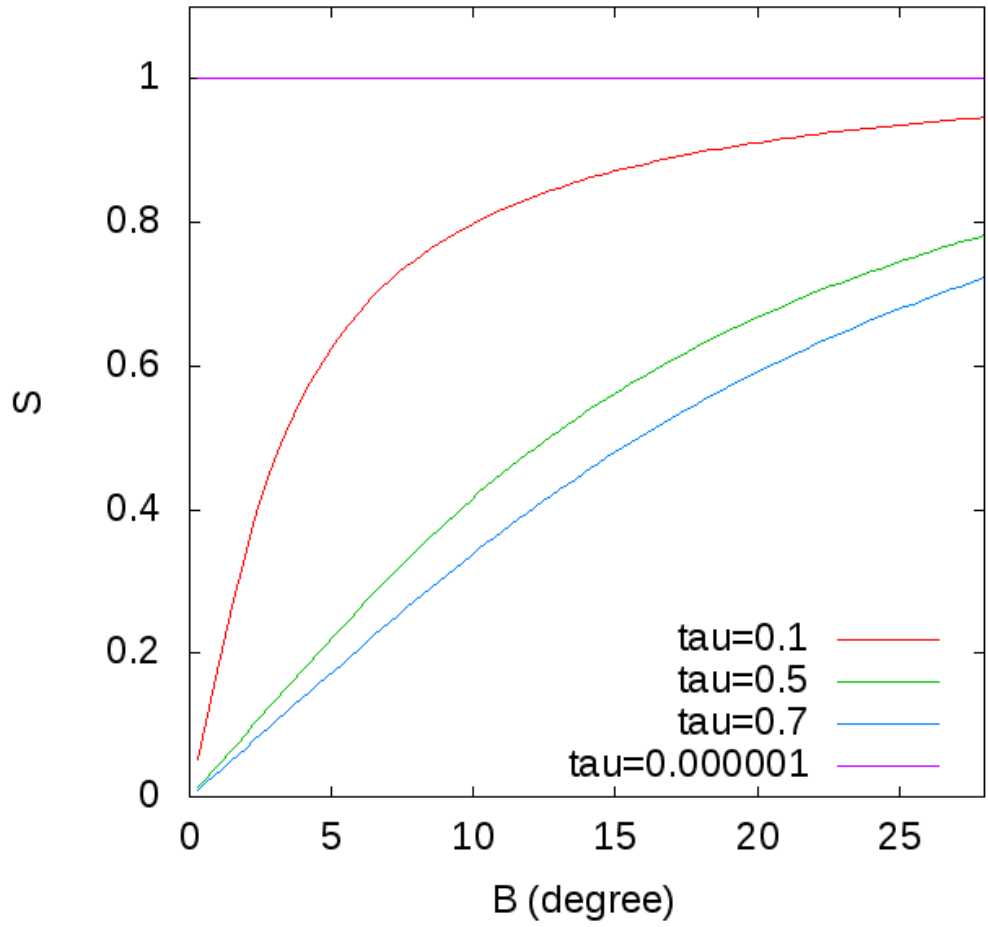

**Figure 4.** Variation in the factor $S$ due to the shadowing of surrounding particles as a function of the ring-opening angle $B$ and the ring optical depth, $\tau$.

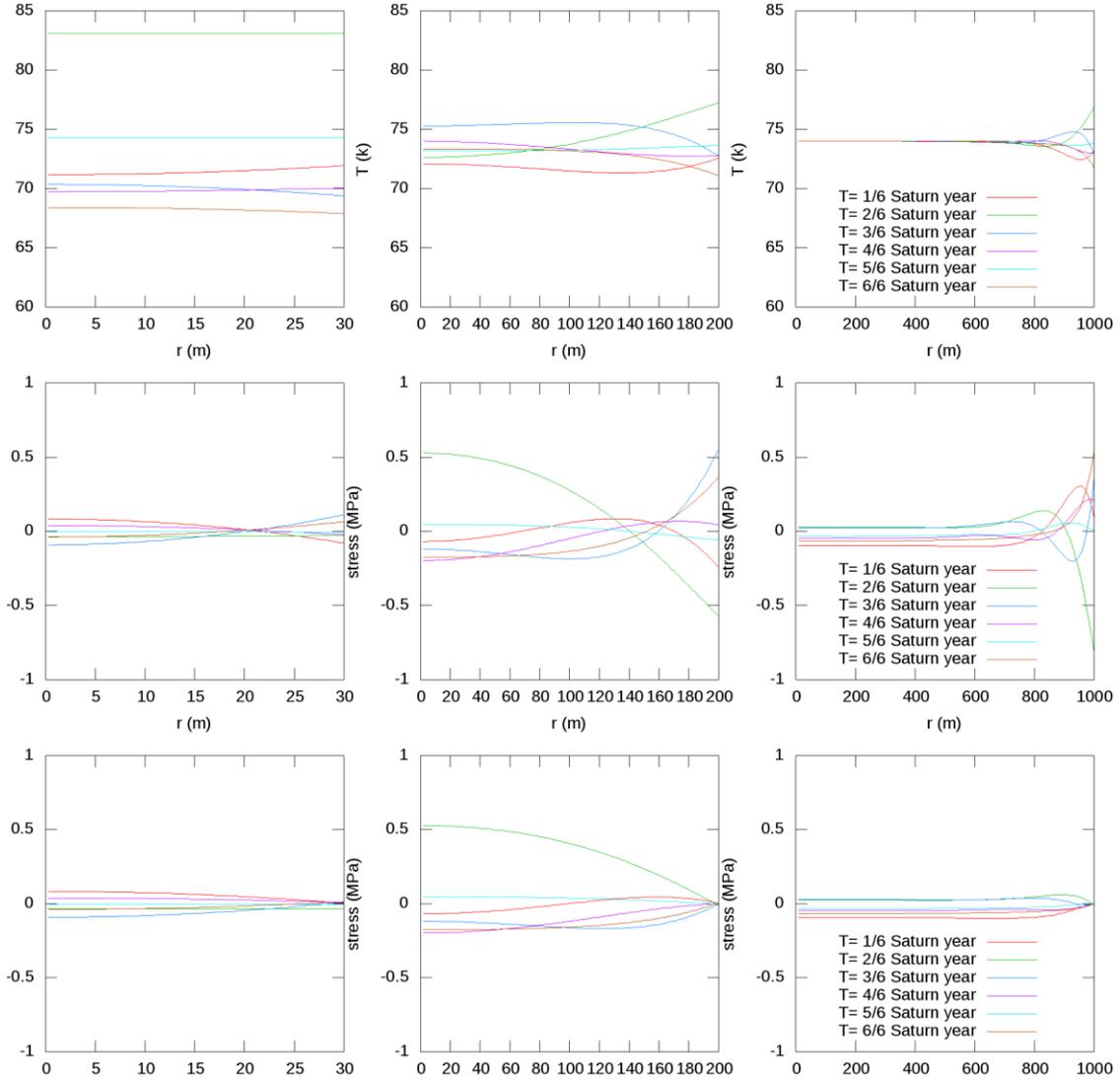

**Figure 5.** Case of non-porous particle floating in the B ring ($\tau \sim 0.7$). The temperature profile (top); tangential stress component, $\sigma_t$ (middle); and radial stress component, $\sigma_r$ (bottom) of a non-porous particle with a radius of $R$ = 30 m (left), 200 m (middle), and 1000 m (right) are shown. In the horizontal axis, $r$ refers to the radial distance from the center. In the vertical axis, the temperature in K (top) and the stress in MPa (middle/bottom) are plotted, where a positive value means that the stress is tensile and negative means it is compressive. Movie of this figure is shown in Movie S1.

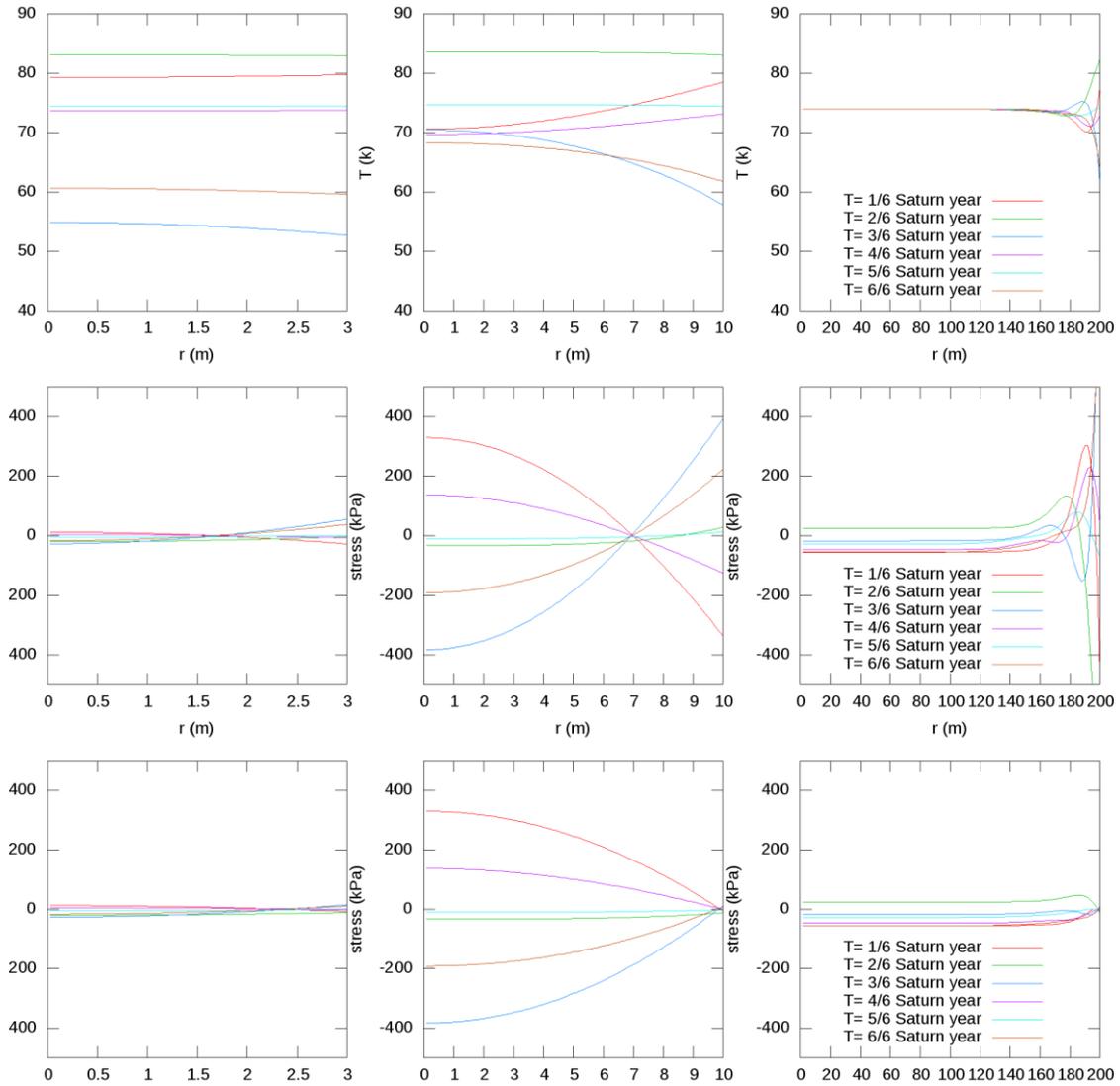

**Figure 6.** Case of a porous particle floating in the B ring ($\tau \sim 0.7$). The temperature profile (top); tangential stress component, $\sigma_t$ (middle); and radial stress component, $\sigma_r$ (bottom) of a highly porous particle with a radius of 3 m (left), 10 m (middle), and 200 m (right) are shown. In the horizontal axis, $r$ refers to the radial distance from the center. In the vertical axis, the temperature in K (top) and the stress in kPa (middle/bottom) are plotted, where a positive value means that the stress is tensile and negative means it is compressive. Movie of this figure is shown in Movie S2.

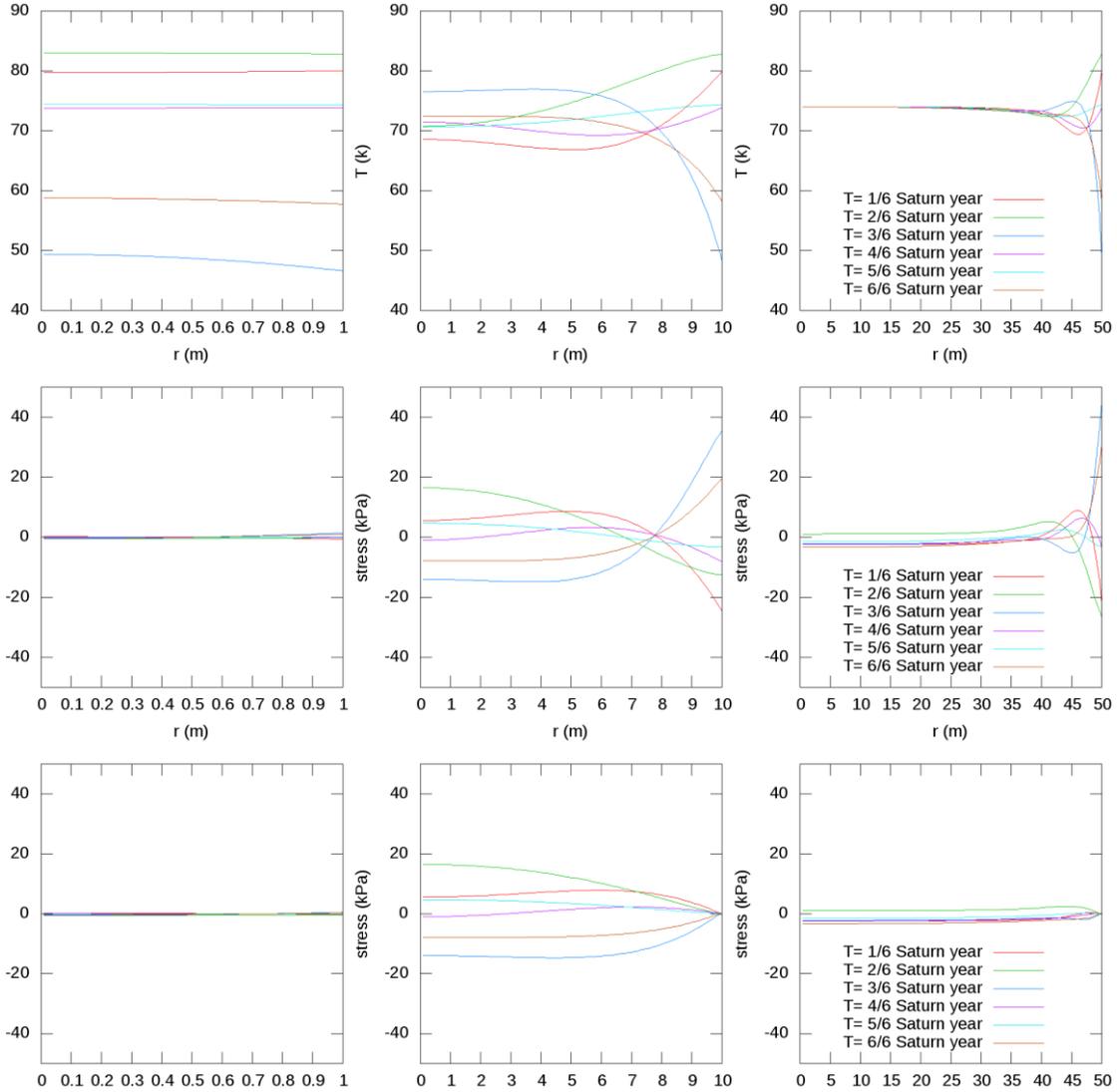

**Figure 7.** Case of a highly porous particle floating in the B ring ($\tau \sim 0.7$). The temperature profile (top); tangential stress component, $\sigma_t$ (middle); and the radial stress component, $\sigma_r$ (bottom) of a highly porous particle with a radius of 1 m (left), 10 m (middle), and 50 m (right) are shown. In the horizontal axis, $r$ refers to the radial distance from the center. In the vertical axis, the temperature in K (top) and the stress in kPa (middle/bottom) are plotted, where a positive value means that the stress is tensile and negative means it is compressive. Movie of this figure is shown in Movie S3.

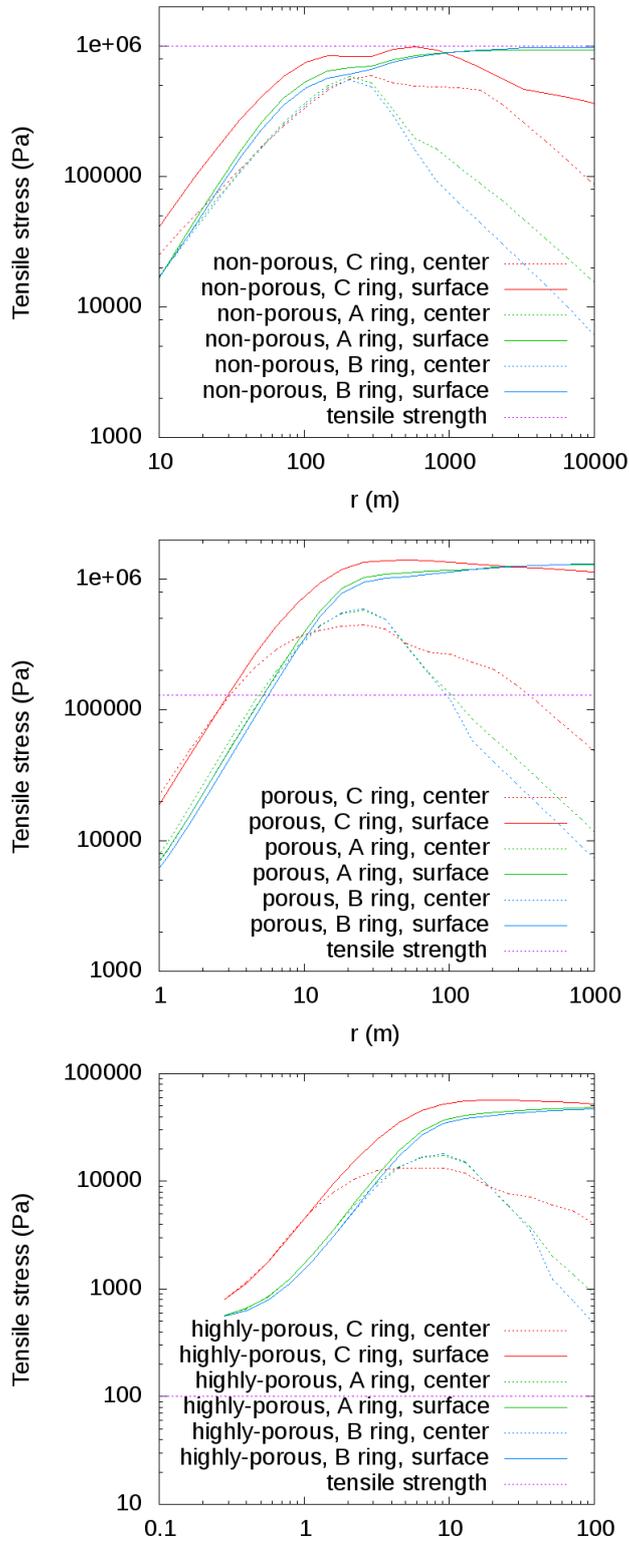

**Figure 8.** Maximum tensile thermal stress in one Saturn year in the center and on the surface of a non-porous (top), porous (middle), and highly porous

particle (bottom). Horizontal brown line reflects the tensile strength of each particle model.

**Appendix A.**

We estimated the short-term temperature changes of the ring particles. As an example, we show the case of the passage of Saturn's shadow; however, other short-term temperature variations are almost equivalent to this case. During the passage of Saturn's shadow (2 h), a ring particle is cooled down due to infrared emission ($\epsilon \sigma T^4$), and its temperature change occurs within the skin depth ($d_\omega$). When the particle size is sufficiently larger than the skin depth ($R \gg d_\omega$), the temperature evolution of the particle surface can be approximated by dividing the thermal emission from the unit area of the particle surface by the heat capacity within the skin depth:

$$\frac{\partial T}{\partial t} = -\frac{\epsilon \sigma T^4}{\rho C d_\omega}, \tag{A1}$$

Assuming the initial temperature of a ring particle to be 100 K, we can obtain the temperature change in the order of $\Delta T = 0.1$ K for the non-porous particle, $\Delta T = 1$ K for the porous particles, and $\Delta T = 10$ K for the highly porous particles. The case of non-porous particles cannot provide sufficient thermal stress to break the particles. The porous and highly porous particles would provide sufficient thermal stress to break the particles, although their skin depths were on the order of 1 mm.

**Acknowledgments**

This work was partly supported by JSPS KAKENHI Grant Numbers 15H03716 (N. H. and K.O.), 20K14538, and 20H04614 (N.H.), 21H00043 (K.O.), and by the Hyogo Science and Technology Association (N.H.). We wish to thank D. Uribe-Suárez and an anonymous reviewer for their comments, which significantly tightened the manuscript.

**Supporting information**

**Movie S1.** Case of non-porous particles floating in the B ring ($\tau = 0.7$). Temperature profile (top) and the tangential and radial stress components, $\sigma_t$ and $\sigma_r$ (bottom) are shown for a non-porous particle with a radius (R) of 30 m (left), 200 m (middle), and 1000 m (right). The horizontal axis, $r$, is the radial distance from the center. The vertical axis indicates the temperature

in K (top) and the stress in MPa, where a positive value indicates that the stress is tensile and a negative value indicates a compressive stress.

**Movie S2.** Case of porous particles floating in the B ring ($\tau = 0.7$). Temperature profile (top) and the tangential and radial stress components, $\sigma_t$ and $\sigma_r$ (bottom) of a non-porous particle with a radius of 1 m (left), 10 m (middle), and 50 m (right) are plotted. in the horizontal axis, $r$ refers to the radial distance from the center. The vertical axis indicates the temperature in K (top) and the stress in kPa, respectively, where a positive value indicates a tensile stress, and a negative value indicates a compressive stress.

**Movie S3.** Case of highly porous particles floating in the B ring ($\tau = 0.7$). Temperature profile (top) and the tangential and radial stress components, $\sigma_t$ and $\sigma_r$ (bottom) of a non-porous particle with a radius of 1 m (left), 10 m (middle), and 50 m (right). in the horizontal axis, $r$ refers to the radial distance from the center. The vertical axis indicates the temperature in K (top) and the stress in kPa, where a positive value reflects a tensile stress, and a negative value indicates a compressive stress.